\documentclass{aastex}      

\usepackage{epsfig}

\newcommand{\omm}{\Omega_{m0}}
\newcommand{\omlam}{\Omega_{\Lambda}}
\newcommand{\omq}{\Omega_{Q0}}

\newcommand{\bc}{\begin{center}}
\newcommand{\ec}{\end{center}}
\newcommand{\rat}{\mathcal R_0}
\begin{document}         
\title{\bf Smooth Energy: Cosmological Constant or Quintessence?}

\author{Sidney Bludman}   
\affil{Deutsches Elektronen-Synchrotron DESY, 22607 Hamburg, Germany\\
University of Pennsylvania, Philadelphia, PA 19104\thanks
{Assisted in part by U.S. Department of Energy grant DE-FGO2-95ER40893.}}
\and \author{Matts Roos, Department of Physics
University of Helsinki, Helsinki, Finland}

\begin{abstract}
  For a flat universe presently dominated by smooth energy, either
  cosmological constant (LCDM) or quintessence (QCDM), we calculate
  the asymptotic collapsed mass fraction as function of the present
  ratio of smooth energy to matter energy $\rat$.  Identifying the
  normalized collapsed fraction as a conditional probability for
  habitable galaxies, we observe that the observed present ratio $\rat
  \sim 2$ is likely in LCDM, but more likely in QCDM. Inverse
  application of Bayes' Theorem makes the Anthropic Principle a
  predictive scientific principle: the data implies that the prior
  probability for $\rat$ must be essentially flat over the
  anthropically allowed
  range. Interpreting this prior as a distribution over {\em theories}
  lets us predict that any future theory of initial conditions must be
  indifferent to $\rat$.  This application of the Anthropic Principle
  does not demand the existence of other universes.

\end{abstract}

\section{A Flat Low-Density Universe}

In the absence of a recognized symmetry principle protecting its
value, no theoretical reason for making the cosmological constant zero
or small has been found.  Inflation makes the universe appear flat, so
that, at present, the vacuum or smooth energy density
$\Omega_{Q0}=1-\omm < 1$, is $10^{120}$ times smaller than would be
expected on current particle theories.  To explain this small but
non-vanishing present value, a dynamic vacuum energy, quintessence,
has been invoked: a background scalar field whose potential energy
dominates its kinetic energy, so that the fluid pressure $P$ and its
ratio to energy density $w_Q \equiv P/\rho <0$.  (When we speak of a
static vacuum energy or cosmological constant, we mean the limiting
case, $w_Q=-1$, which is homogeneous on all scales.)  With any
positive cosmological constant or quintessence, an expanding universe
starts out radiation or matter dominated, but ultimately becomes
dominated by smooth energy and goes into exponential expansion (Fig.
1).
   
The best evidence for a flat low-density universe comes from the
location (at $l \sim 200$) of the first Doppler peak in the CBR
anisotroy in the combined BOOMERANG-98, MAXIMA-1 and COBE-DMR
measurements: $\omm+\omq=1.11 \pm 0.07~^{+0.13}_{-0.12}$
\citep{Jaffe}.  Supporting evidence \citep{WCOS,RHOR} comes from the
slow evolution of rich clusters, the mass power spectrum, the
curvature in the SNIa Hubble diagram, and the dynamic age.  The cosmic
flow implies $\omm=0.3\pm0.05$.  The height of the first Doppler peak,
and gravitational lensing imply $\omq=1-\omm \sim 2/3$.  Of these, the
SNIa evidence is most subject to systematic errors due to precursor
intrinsic evolution.

A large set of such observational data have been combined \citep{RHOR} in a
two-step constrained fit. Firstly, ten independent constraints in
the ($\omm,\omlam$)-plane yielded the result $\omm + \omlam = 0.99 \pm
0.14$, which clearly supports the view of a flat universe. Secondly,
assuming exact flatness, five more constraints were included in the
fit with the result $\omm = 1 - \omlam = 0.33 \pm 0.04$, or
equivalently, $\rat=\omq/\omm = 2.03\pm 0.25$. We can interpret this
as evidence that we live in a low-density universe with a smooth
energy component with present density $\omq \sim 2/3$ and negative
pressure $ -1 \leq w_Q < -1/3 $.

Accepting this small but non-vanishing value for static or dynamic
vacuum energy, a flat Friedmann cosmology (CDM) is characterized by
$\omm,~\Omega_{Q0}=1-\omm$ or the present ratio
 $$\rat \equiv \Omega_{Q0}/\omm=(1-\omm)/\omm~, $$
and by the equation of state for the smooth energy component.
The {\em Cosmic Coincidence} problem now becomes
pressing: Why do we live when the clustered matter density
$\Omega(a)$, which is diluting as $a^{-3}$ with cosmic scale $a$, is
just now comparable to the static vacuum energy or present value of
the smooth energy i.e. when the ratio $\rat \sim 2$ ?  
 
In this paper, we study the {quintessence} range $ -1 \le w_Q < -1/3 $
for the smooth energy component, distinguishing in particular the two
cases \\ LCDM: cosmological constant with $w_Q=-1$, and\\ QCDM:
quintessence with the specific choice $ w_Q =-1/2$ . 

The next section compares the cosmic expansion and the freeze-out of
structure formation, in these two models for smooth energy. Section 3
extends to QCDM the calculation of asymptotic mass fraction as
function of a hypothetical continuous variable $\omm$ for LCDM,
presented by Martel {\it et al} \citep{MSW,MS}.  In Section 4,
identifying these collapsed mass fractions with anthropic
probabilities for $\rat$, we show that the presently observed ratio,
while reasonable in an LCDM universe, is more likely in a QCDM
universe.  This confirms empirically that the prior probability for
our universe is flat in $\omm$, as is expected in a large class of
fundamental theories \citep{W}.
 
The anthropic answer to the cosmic coincidence problem is:
``If not now, then when?'' \citep{Hil}

\section {Expansion of a Low Density Flat Universe}

The Friedmann equation in a flat universe with clustered matter and
smooth energy density is
$$
H^2(x) \equiv (\dot{a}/a)^2=(8 \pi G/3)(\rho_m+\rho_Q), $$
or, in units of $\rho_{cr}(x)=3H^2(x)/8\pi G$,
$$1=\Omega_m(x)+\Omega_Q(x),$$
where the reciprocal scale factor $x
\equiv a_0/a \equiv 1+z \rightarrow \infty$ in the far past,
$\rightarrow 0$ in the far future.

With the effective equation of state $w \equiv P/\rho=$ constant,
different kinds of energy density dilute at different rates $\rho \sim
a^{-n},~n \equiv 3(1+w)$, and contribute to
the deceleration at different rates $(1+3w)/2$ shown in the table:\\
   \begin{table}
   \centering
   \begin{tabular*}{115mm}{@{\extracolsep{\fill}}l|ccc@{}}  \hline
   
   {\em substance}              &{\em w}       &{\em n}       &{\em (1+3w)/2}   \\  \hline
   radiation      &  1/3   &  4     &1    \\
   NR matter      &   0    &  3     &1/2      \\
   quintessence   & -1/2   &  3/2   &-1/4   \\
   cosmolconst    & -1     &  0     &-1     \\
   \hline  
   \end{tabular*}
   \caption{Energy Dilution for Various Equations of
       State}
   \end{table}\\
   The expansion rate in present Hubble units is
$$ H(x)/H_0=(\omm x^3+ (1-\omm) x^n_Q)^{1/2}. $$
The Friedmann equation has an unstable fixed point
in the far past and a stable attractor in the far future.  (Note the
tacit application of the anthropic principle: Why does our universe
expand, rather than contract?)

\begin{figure}
\begin{center} 
\epsfig{file=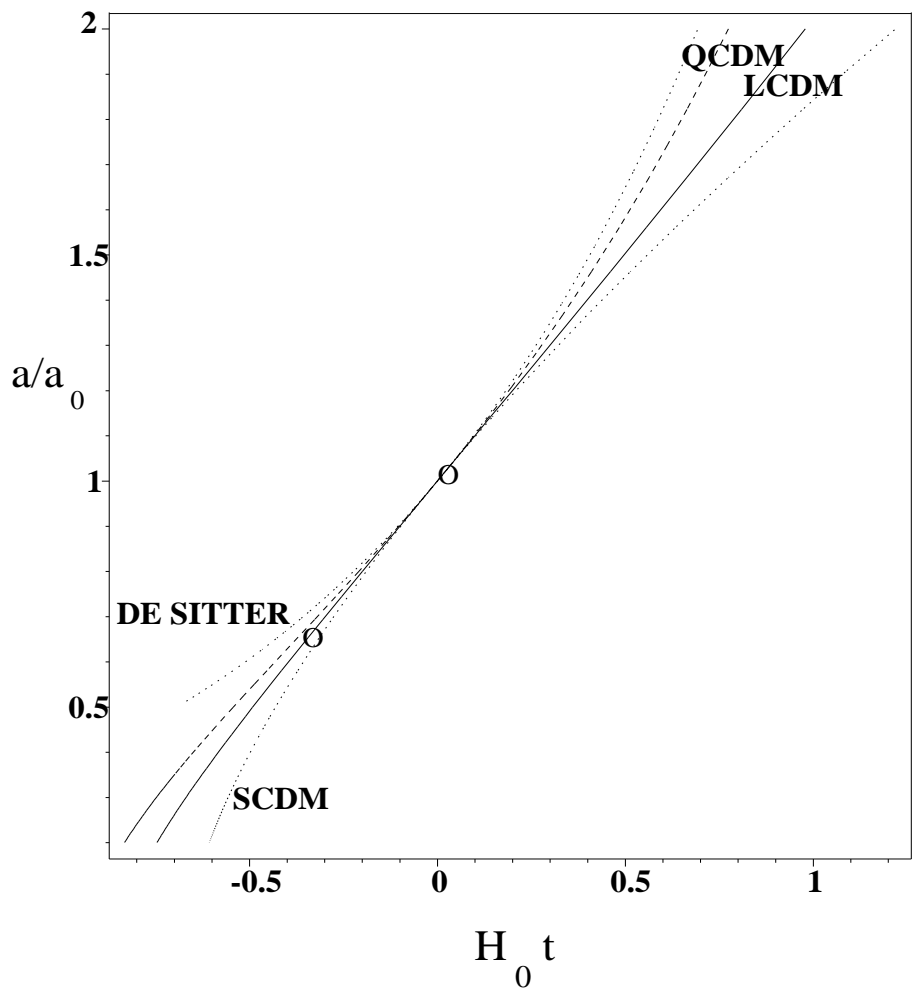,width=12cm,height=13cm} 
\caption{Scale evolution of LCDM and QCDM low-density flat universes
in the recent past and near future. The lower curve shows
the SCDM universe from which both LCDM and QCDM evolved in
the far past. The upper curve shows
the flat de Sitter universe towards which both LCDM and QCDM will evolve in
the future. The inflection points marked (O) show where first LCDM
and later QCDM change over from decelerating to accelerating universes.}  
\end{center}
\end{figure}
\nopagebreak
The second Friedmann equation is 
$$q(a)\equiv -\ddot{a} a/{\dot{a}^2}=\sum_{i} \Omega_i (1+3w_i)/2=
(1+\Omega_Q(a))/2 .$$ 
The ratio of smooth energy to matter energy, $\mathcal R(a)=\mathcal
R_0 (a_0/a)^{3w_Q}$, 
increases as the cosmic expansion dilutes the matter density.
A flat universe, characterized by $\rat,~w_Q$, evolves out of an SCDM
universe in the remote past towards a flat de Sitter universe in the future.  
As shown by the inflection points (O) on the middle curves of Figure 1,
for fixed $\rat$, QCDM  expands
faster than LCDM, but begins accelerating only
at the present epoch.  The top and bottom curves refer respectively to
a de Sitter universe ($\Omega_m=0$), which is always accelerating, and
an SCDM universe ($\Omega_m=1$), which is always decelerating.

The matter-smooth energy transition (``freeze-out'')
$\Omega_Q/\Omega_m=1$ took place only recently at $(x^*)^{-w_Q}=\rat^{-1/3}$ 
or at $x^*=\rat^{2/3}=1.59$ for QCDM and, even
later, at $x^*=1.26$ for LCDM. Because, for the same value of
$\rat$, a matter-QCDM transition would take place earlier and more
slowly than a matter-LCDM transition, it imposes a stronger constraint
on structure evolution. 
As summarized in the table below, quintessence dominance begins 3.6
 Gyr earlier and more gradually than cosmological constant dominance.
 (In this table, the deceleration $q(x) \equiv -\ddot{a}/aH_0^2$ is
 measured in {\em present} Hubble units.)  The recent lookback time is
$$H_0t_L(z)=z-(1+q_0)z^2+...,\quad z<1 ,$$
where $q_0=0$ for QCDM and $q_0=- 1/2$ for LCDM.

   \begin{table} 
   \centering
   \noindent
   \begin{tabular*}{125mm}{@{\extracolsep{\fill}}l|cc@{}}  \hline
   {\em event}                 &  {\em LCDM}      &  {\em QCDM} \\
   \hline \hline
   {\bf Cross-Over to Smooth Energy Dominance}&               &             \\
   reciprocal scale $x^*=a_0/a=1+z$                &$\rat^{1/3}$=1.260           &$\rat ^{2/3}$=1.587     \\
   age   $t(x^*)/H_0^{-1}$    &0.720             &0.478           \\
    in units $h_{65}^{-1}$Gyr      &10.8              &7.2            \\ \hline
   horizon size in units $cH_0^{-1}$   &2.39              &1.58           \\
    in units $h_{65}^{-1}$Gpc      &11.0              &7.24            \\ \hline
   deceleration $q(x^*)$ at freeze-out               &-0.333            &0.333
   \\  \hline \hline
   {\bf Present Epoch}         &                  &              \\
   age $t_0/H_0^{-1}$         &0.936             &0.845          \\
         $h_{65}^{-1}$Gyr      &14.0              &12.7           \\ \hline
   horizon in units $cH_0^{-1}$                      &3.26              &2.96          \\
    in units $h_{65}^{-1}$Gpc &15.0              &13.6            \\ \hline
   present deceleration $q_0$  &-0.500            &0              \\ \hline

   \end{tabular*}
   \caption{ Comparative Evolution of LCDM and QCDM Universes}
   \end{table}

\section{Evolution of Large Scale Structure}

In this section, we extend to QCDM earlier LCDM calculations
\citep{MSW,GLV,MS} of the asymptotic mass fraction $f_{c,\infty}$ that
ultimately collapses into evolved galaxies.  This is presumably a
measure of the number density of galaxies like our own, that are
potentially habitable by intelligent life. 
We then compare the QCDM and LCDM asymptotic
mass fraction distribution functions, as function of an assumed
$\omm$.

The background density for large-scale structure formation
is overwhelmingly Cold Dark Matter (CDM), consisting of clustered
matter $\Omega_m$ and smooth energy or quintessence $\Omega_Q$.
Baryons, contributing only a fraction to $\Omega_m$, collapse after
the CDM and, particularly in small systems, produce the large
overdensities that we see.
  
Structure formation begins and ends with matter dominance, and is
characterized by two scales: The horizon scale at the first
cross-over, from radiation to matter dominance, determines the power
spectrum $P(k,a)$, which is presently characterized by a shape factor
$\Gamma_0=\omm h =0.25 \pm 0.05$.  The horizon scale at the second
cross-over, from matter to smooth energy, determines a second scale
factor, which for $w_Q=-1/2$ quintessence, is at $\sim 130$ Mpc, the scale of
voids and superclusters.  A cosmological constant is smooth at all
scales.
  
Quasars formed as far back as $z \sim 5$, galaxies at $z \geq 6.7$,
ionizing sources at $z=(10-30)$.  The formation of {\em any} such
structures, already sets a large upper bound $x^*<30$ or $(\omlam /
\omm)<1000, \omq<30$, for {\em any} structure to have formed.  A much
stronger upper bound, $u_0<5$, is set by when {\em typical} galaxies
form i.e. by estimating the {\em probability} of our observing
$\rat=2$ at the present epoch.

\subsection{Asymptotic Collapsed Mass Parameter $\beta$}  
Garriga {\it et al} \citep{GLV} and Bludman \citep{Blud} have
already calculated the asymptotic mass fraction from the
Press-Schechter formalism
$$f_{c,\infty}(\beta) =\mbox{erfc}( \sqrt\beta
)=(2/\sqrt\pi)\int^{\infty}_{\sqrt\beta}\exp(-t^2)\,dt , $$
depending only on
$$\beta \equiv \delta_{i,c}^2/(2\sigma_i^2) , $$
where $\sigma_i^2$ is
the variance of the density field, smoothed on some scale $R_G$, and
$\delta_{i,c}$ is the minimum density contrast at recombination which
will ultimately make a bound structure.  This minimum density contrast
grows with scale factor $a$, and is, except for a numerical factor of
order unity (Eq.(\ref{const}\citep{MS}), $\delta_{i,c} \sim
x^*/(1+z_i)$.  Both numerator and denominator in $\beta$ refer to the
epoch of recombination, but this factor $(1+z_i)$ cancels out in the
quotient.

MSW \citep{MSW} and MS \citep{MS} have
improved on the Press-Schechter formalism by assuming spherical
collapse of Gaussian fluctuations or linear fluctuations that are
surrounded by equal volumes of compensating underdensity.  Except in
the limit $\beta \rightarrow 0$, the PS formula overestimates the
collapsed mass by factor $\approx (1.70)\cdot\beta^{0.085}$, or about 50\%
near $\omm=1/3$ \citep{Blud}. Here we will
use the improved MSW formula for both $R_G=1,~2$ Mpc,
\begin{eqnarray}
f_{c,\infty}(\beta)={1\over \sqrt \pi}\int^{\infty}_{\beta}{{\exp(-x) dx}\over 
{\sqrt x + \sqrt\beta}}\ . 
\end{eqnarray} 

The variance of the mass power spectrum depends on the cosmological
model ($\omm$) and on the relevant co-moving galactic size scale
$R_G$, but is insensitive to $w_Q$, for $w_Q<-1/3$ \citep{WS}.  For the
QCDM model we consider, $\sigma_i^2(\omm,R_G)$ is therefore the same
as that already calculated \citep{MSW,MS} for LCDM, for a
scale-invariant mass spectrum smoothed with a top-hat window function.
For the observed ratio $\rat=2,~\omm=1/3$, the value of $\sigma_i
\cdot 10^{-3}$ at recombination is 3.5 and 2.4 for comoving
galactic size scale $R_G=1$, and 2 Mpc, respectively.

For a flat universe the numerical factor in $\delta_{i,c}$ is given by
\citep{MS}
\begin{eqnarray}
{3\over 5}{(3-n)\over {(2-n)^{(2-n)/(3-n)}}}~~ ,\quad n=3(1+w_Q).
\label{const}\end{eqnarray}
Thus $\delta_{i,c}=1.1339 x^*/(1+z_i)$ for both $n=0$ and $n=3/2$. The
collapsed mass parameter $\beta=(1.1339/2)\cdot
[x^*/\sigma_i(R_G,\rat)]^2$, depends explicitly on $\rat$ for LCDM and
QCDM.  It also depends implicitly on $\rat$ through $\sigma_i$.
Nevertheless, in going from LCDM to $w_Q=-1/2$ QCDM, the argument of
$f_{c,\infty}$ scales simply as
$\beta_{QCDM}=\beta_{LCDM}\cdot\rat^{1/3}$.

Both asymptotic mass fractions are practically unity for large $\omm$,
but fall off with increasing ratio $\rat >1$.  For any $\rat>1$ or
$\omm < 0.5$, QCDM always leads to a smaller asymptotic mass fraction
than LCDM. For ratio $\rat < 1$, $f_{c,\infty}$ changes slowly and the
differences between QCDM and LCDM are not large.

\subsection{Asymptotic Collapsed Mass Fraction Distribution Function}
As function of the ratio $\rat$, the asymptotic mass fraction defines
a distribution function
$$f_{c,\infty}=d\mathcal P/d \rat .$$
In Figure 2, instead
of $f_{c,\infty}$ we plot the logarithmic distribution function in the
ratio $\rat$
\begin{eqnarray}
F(\omm)=\rat \cdot f_{c,\infty}=d \mathcal P/d \log \rat , 
\label{F}\end{eqnarray}
for
LCDM and for QCDM and galactic size scale 1 Mpc. (Even for LCDM, this
differs by a factor $\sigma_i ^3(\omm)$ from the logarithmic
distribution in $\beta$, $d\mathcal P/d \log (\beta^{3/2})$ that is
plotted in \cite{MSW} and \cite{GLV}.) $F(\omm)$ may be thought of as the ratio
$\rat$ weighted by the number density of galaxies $f_{c,\infty}$.

Figure 2 shows broad peaks in the logarithmic distributions in $\omm$
at $(\omm,F,R_G)=(0.23,1.09,1$ Mpc) and $(0.28,0.78,2$ Mpc) for LCDM,
and at $(0.32,0.80,1$ Mpc) and $(0.37,0.61, 2$ Mpc) for QCDM. At the
observed $\omm=1/3$, the LCDM asymptotic mass fraction logarithmic
distributions in $\rat$ fall 13\%, 5\% below the LCDM peaks for $R_G=$
1,2 Mpc respectively. The QCDM distributions are peaked nearer
$\omm=1/3$, and peak only 0.3\%, 1.7\% below the QCDM peaks for 
$R_G=$ 1, 2 Mpc respectively. These asymptotic collapsed mass curves
have not yet been normalized to unit area.

In order to interpret these distributions as differential
probabilities $\mathcal P(\omm)$, we now normalize the function
$F(\omm)$ in Eq.(\ref{F}, by dividing the $R_G=1$ Mpc curves by 0.378,
0.485  for QCDM, LCDM respectively and the $R_G=1$ Mpc curves by 0.289,
0.364 respectively.  In Figure 3, we plot this conditional
probability at $\rat=2$ as a function of $w_Q$ for $R_G=1$ and 2
Mpc. At every $w_Q$, $\rat=2$ is more probable in QCDM than in LCDM,
particularly for the smaller galactic mass smoothing scale. For $R_G=$
1 Mpc, $w_Q=-1/2$ QCDM is 10\% more probable than LCDM.

It is not surprising that our universe, containing at least one
habitable galaxy, has $\rat=\mathcal O(1)$. What is impressive is that
our observed low-density universe, is almost exactly that which will
maximize the number density of habitable galaxies.  Our existence does
not explain $\omm$, but the observed value makes our existence (and
that of other evolved galaxies) most likely.

\begin{figure}
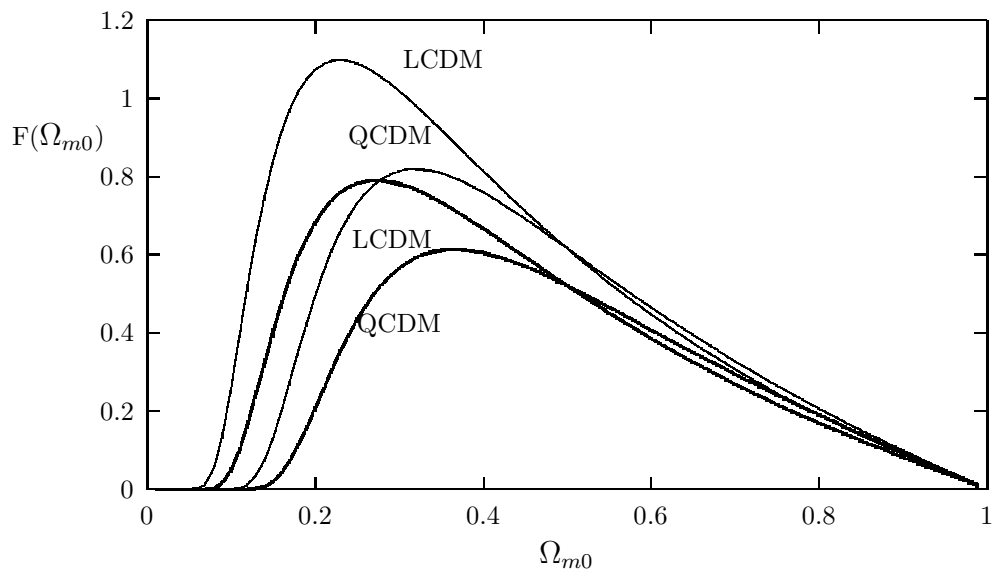

\begin{center}  
\input figq3.tex
\caption{Logarithmic distribution function for the asymptotic
collapsed mass fraction as function of hypothetical present matter
density $\omm$. The smoothing scale is taken to be $R_G=1$ Mpc for the
thin line curves, and 2 Mpc for the thick line curves. Our observed
universe with $\omm \sim 1/3,~ \rat \sim 2$ falls within the broad
peak of the LCDM distributions and close to the peak of the
QCDM distributions.}
\end{center}     
\end{figure} 
\begin{figure}
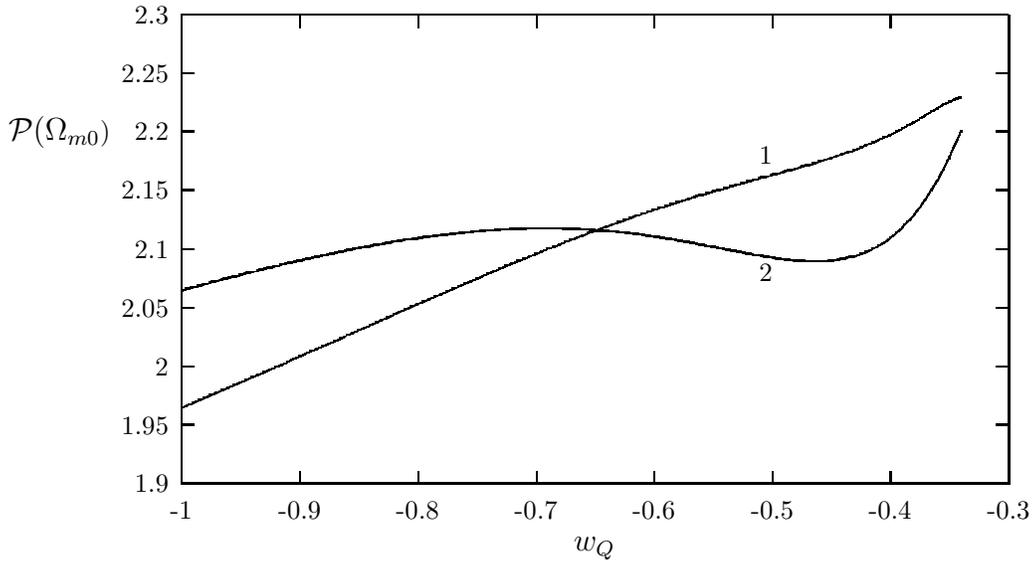

\input figq2.tex
\begin{center}  
\caption{The effect of quintessence equation of state on the
  asymptotic mass fraction logarithmic distribution function for the observed
  ratio $\rat=2$.  The normalized logarithmic distribution function is shown as
  function of $w_Q$ for each of the two smoothing scales $R_G=1,\ 2$
  Mpc. Particularly for $R_G=$ 1 Mpc, the observed ratio is
  more likely in QCDM than in LCDM.} 
\end{center}     
\end{figure}

\section{$\Omega_{m0}\sim 1/3$ is Quite Likely  for Our Universe}
\subsection{The Datum Implies that the Prior is Flat}
What inference should we draw
from the datum that our observed universe lies at or near the
peak in the asymptotic mass fraction logarithmic distributions for a
QCDM or LCDM universe?

Bayes' Theorem  makes the posterior probability $\mathcal
P_{post}(\omm)$ of observing a particular value of $\omm$ 
\begin{eqnarray}
\mathcal P_{post}(\omm)=\mathcal P(\omm)\cdot \mathcal P_{prior}(\omm)\ ,
\end{eqnarray}
where $\mathcal P(\omm)$ is $F(\omm)$ divided by the normalization
factors given in Sec. 3.2.  The posterior probability always depends
directly on the assumed prior $\mathcal P_{prior}(\omm)$, which
measures our subjective hypotheses about $\omm$, and should ultimately
be determined by the initial conditions. From the fact that our
universe falls at or near the peak of the logarithmic asymptotic mass
distribution, we can infer that the prior for the logarithmic distribution
\begin{eqnarray}
\mathcal P_{prior}(\omm)=\mathcal P_{post}(\omm)/\mathcal P(\omm)\ .
\end{eqnarray}
is flat, at least for $\rat=\mathcal O(1)$.  (Alternatively, if we
chose $\mathcal P(\omm)$ to be simply the asymptotic mass fraction
$f_{c,\infty}$, we would infer that the prior
must be proportional to $\rat$. Quasi-philosophical arguments
depending on ``simplicity'' or Occam's Razor \citep{Jeff}
favor defining $\mathcal P(\omm)$ by the logarithmic distribution,
rather than by such a linear distribution, particularly when the data
is sparse.)

Garriga and Vilenkin \citep{GV} argue that, for many theories, the
prior is {\em not} flat.  MSW, assuming nothing about initial
conditions, assume a prior flat in $\omm$.  Indeed, \citep{W} finds
that the prior will be flat for a large class of theories: those with
slow-roll potential $V(\phi)=V_1 f(\lambda\phi)$ where $V_1 \sim
\mathcal O(M_P^4)$ is a large
energy density, $f(x)$ is a dimensionless function involving no very
large or very small parameters, and $\lambda$ is a very small
dimensional parameter.

\subsection{Anthropic Interpretation of the Flat Prior}
The Anthropic Principle asserts that the probability of habitable
galaxies, solar systems and intelligent observers is proportional
to the posterior probability $\mathcal
P_{post}(\omm)$.
The datum $\omm \sim 1/3$, at or near the peak of the logarithmic
asymptotic mass distribution, then infers only that its prior is flat, i.e.
is indifferent to hypotheses concerning $\omm$, at least for
$\rat=\mathcal O(1)$. It is essential to recognize that the prior is a
functional of {\em hypotheses}, not of the data sample \citep{Lor}.  What is
the observation $\rat \sim 2$ telling us that we didn't believe before-hand?

A popular anthropic interpretation has already been given
\citep{E,V,MSW,GLV} to ``explain'' the observed non-vanishing
cosmological constant.  These authors take seriously a meta-universe
containing an infinite ensemble of real subuniverses with all possible
values for the vacuum energy $\omlam=1-\omm$.  In each of these
subuniverses, $\omlam$ determines $\mathcal P(\omm)$, the normalized
probability for habitable galaxies to have emerged before the present
epoch.  Our habitable subuniverse is rare, only one of many more
subuniverses with inhospitable values for $\omm$.  Many theories of
cosmology and of quantum mechanics do predict other sub-universes,
with different values of the fundamental constants, or even of the
physical laws. Nevertheless, some of these other sub-universes need to
be ultimately observable by us, at least in principle, if this
VAnthropic Principle is to be a falsifiable physical theory.  This
many-world interpretation of the flat prior is suggested by eternal
inflation and by quantum cosmology, but close to a frequentist
probability view and practically impossible to test. It is not required by
the data.

Indeed, the present situation in cosmology is an ideal case for the
proper use of Bayes' Theorem: in the face of (presently) incomplete
information, to make statistical inference concerning {\em
hypotheses}. A modest inverse probability application of Bayes'
Theorem does not require a present distribution of potentially
observable subuniverses.  (Our own universe might, of course,
ultimately evolve from or towards different universes with different
values for the cosmological parameters.)  Instead, our calculation of
the flat prior merely asserts that our knowledge of $\omm$ is
consistent with those slow-roll scalar field theories \citep{W}
which are indifferent to $\omm$. 

In summary, our anthropic interpretation of the ratio of smooth energy
to clustered mass in our own universe, $\rat \sim 2$, predicts a large
class of generic quintessence models and slightly prefers $w_Q=-1/2$
QCDM over LCDM.
 
This research benefited from useful discussions with H. Martel.

\end{document}